\documentstyle[epsf,psfrag,11pt]{preprint}
\input amssym.def
\input amssym.tex
\font\bbb=msbm10 scaled 1100

\newtheorem{thm}{Theorem}[section]
\newtheorem{lemma}[thm]{Lemma}
\newtheorem{cor}[thm]{Corollary}
\newtheorem{prop}[thm]{Proposition}

\newtheorem{dfn}[thm]{Definition}

\newcommand{\La}{\Lambda}

\newcommand{\del}{\partial}

\newcommand{\ra}{\rightarrow}
\newcommand{\hra}{\hookrightarrow}
\newcommand{\real}{\mbox{\bbb R}} 	% {\mbox{\rm I} \hspace{-.03in} {\bf R}}
\newcommand{\zed}{\mbox{\bbb Z}}	%{\mbox{\bf Z}\hspace{-.06in}{\bf Z}}
\newcommand{\curl}{\nabla\times}

\newcommand{\Lie}{{\cal L}}

\newcommand{\rest}[2]{\left. #1\right\vert_{#2}}		
\newcommand{\be}{\begin{equation}}
\newcommand{\ee}{\end{equation}}
\newcommand{\bea}{\begin{eqnarray}}
\newcommand{\eea}{\end{eqnarray}}
\newcommand{\bmini}{\footnotesize\begin{center}\begin{minipage}{5.5in}}
\newcommand{\emini}{\end{minipage}\end{center}\normalsize}
\newcommand{\pf}{{\em Proof: }}

\newcommand{\qed}{\hfill$\Box$}

\renewcommand{\L}{{\cal L}}

\newcommand{\T}{{\cal T}}

\newcommand{\V}{{\cal V}}

\newcommand{\eg}{{\em e.g.}}
\newcommand{\ie}{{\em i.e.}}
\newcommand{\cf}{{\em cf. }}
\newcommand{\mathspace}{\;\;\;\;\;\;\;}
\begin{document}
%
% TITLE
%
\vspace{0.35in}

\begin{center}
\large 
{\bf CONTACT TOPOLOGY AND HYDRODYNAMICS III: KNOTTED FLOWLINES}
\normalsize
\vspace{0.5in}

%
% AUTHORS
%
\begin{tabular}{cc}
JOHN ETNYRE & ROBERT GHRIST \\ 
Department of Mathematics & School of Mathematics \\
Stanford University & Georgia Institute of Technology \\
Stanford, CA 94305 & Atlanta, GA 30332-0160
\end{tabular}
%JOHN ETNYRE \hspace{0.2in} ROBERT GHRIST  

\vspace{0.5in}

\end{center}
% 
% ABSTRACT
%
\begin{abstract}
We employ the relationship between contact structures and Beltrami fields 
derived in part I of this series to construct steady nonsingular solutions 
to the Euler equations on a Riemannian $S^3$ whose flowlines trace out closed
curves of all possible knot and link types simultaneously. Using careful
contact-topological controls, we can make such vector fields 
real-analytic and transverse to the tight contact structure on $S^3$.
\vspace{0.1in}

%
% AMS NUMBERS
%
\noindent
{\sc AMS classification: 57M25, 76C05, 58F25, 58F05.}
\end{abstract}

\newpage
% **************************************
\section{Introduction}
% **************************************

This work considers the paths of particles in a three-dimensional 
steady inviscid fluid flow. The relationship between Lagrangian 
dynamics and knot theory comes from the observation that any 
particle path which is periodic traces
out a simple closed curve --- a {\em knot}. We note the historical 
significance of this relationship: Lord Kelvin's notion that matter 
was built from knotted vortex tubes in the \ae ther initiated 
his and Tait's investigations into knot types, in the hope that
the structure of the periodic table would be recovered \cite{Tho69}.
Thus the mathematical classification of knot types began, arguably, 
as a problem in fluid dynamics. 

We consider solutions to the Euler equations for a perfect 
incompressible fluid, restricting attention to steady flows of 
high regularity for simplicity. 
In a previous paper, 
we showed that unknotted orbits are 
forced in steady real-analytic Euler flows on $S^3$.
\vspace{0.1in}

\noindent
{\bf Theorem (\cite{EG:unknot})}{\em 
Any steady $C^\omega$ Euler flow on a Riemannian $S^3$ must possess 
a closed flowline which bounds an embedded disc: an unknot.
}
\vspace{0.1in}

In this work, we consider the opposite end of the spectrum, asking 
``What is possible'' rather than ``What is forced?'' The result 
of our inquiry is that the most complicated and intricate 
knotting and linking phenomena known are present within the simplest 
class of fluid flows.
\vspace{0.1in}

\noindent
{\bf Main Theorem }{\em 
There exists a steady nonsingular $C^\omega$ solution to the 
Euler equations on a Riemannian $S^3$ which possesses 
periodic flowlines of all possible knot and link types simultaneously.
}
\vspace{0.1in}

This theorem answers a question in \cite{Wil98}. General vector fields on 
$S^3$ having all knots and links as closed orbits were discovered by 
Ghrist \cite{G97TOP,GHS97}, using the {\em template}
construction of Birman and Williams \cite{BW83a,BW83b}. 
It is by no means clear that such phenomena can arise within 
fluid flows --- indeed, a large class of solutions (the integrable fields)
of necessity possess a very restricted class of knot types. 
We are thus forced to consider the contact geometry and topology
associated to Euler flows, as elucidated in part I of this series 
\cite{EG:I}. We translate the problem of constructing topologically
complicated Euler flows to the problem of finding a 
certain kind of contact form on $S^3$. A careful 
construction yields a $C^\omega$ solution which is 
furthermore transverse to the standard tight contact structure
on $S^3$ (see \S\ref{sec_Contact} for definitions). 

The advantage of working in the realm of contact topology is that
it is genuinely topological: one may perform surgery 
or cut-and-paste constructions on  
contact forms, and still have a solution to the Euler equations
at the end of the day. However, the price paid is a geometric 
one --- the standard Riemannian structure is necessarily altered by our
constructions. Thus the Euler flows we construct satisfy the 
Euler equations for {\em some} Riemannian structure almost 
certainly different than the standard one. It remains an open 
(and interesting) problem to find such a knotted steady flow
on Euclidean $\real^3$ (or to find an obstruction).

In many respects, this paper is inspired by the pioneering work 
of Moffatt, who, in a series of papers \cite{Mof85,Mof86} discusses 
Euler flows ``with arbitrarily complex topology.'' What is meant
by this is the construction of  steady solutions to 
the Euler equations on Euclidean 
$\real^3$ which realize the same orbit topology as any 
given volume-preserving flow on the space. These results have the
advantage of staying within the class of Euclidean metrics. 
However, there are two caveats associated to this work: (1) the 
techniques do not guarantee a continuous solution --- so-called
``vortex sheet'' discontinuities may develop; (2) the proof 
itself relies crucially 
upon the global-time existence of solutions to the Navier-Stokes 
equations (with an alternate viscosity term). Such an existence 
theorem is to date unknown.

% **************************************
\section{Background}
% **************************************

Since the results of this paper require techniques and perspectives 
from a variety of otherwise disjoint subjects, we include a 
substantial amount of background material. The expert reader 
may skip the following subsections as appropriate. 

% #!#!#!#!#!#!#!#!#!#!#!#!#!#!#!#!#!#!#!#!#!#!#!#!#!#!#!#!#!
\subsection{Beltrami fields}
\label{sec_Beltrami}
% #!#!#!#!#!#!#!#!#!#!#!#!#!#!#!#!#!#!#!#!#!#!#!#!#!#!#!#!#!

For information on a topological approach to the relevant 
equations of fluid dynamics, see the recent monograph \cite{AK97}.

The fundamental equations describing the velocity field $u$ of a 
perfect incompressible fluid flow on a Riemannian three-manifold
$M$ with metric $g$ and distinguished volume form $\mu$ are the 
Euler equations. We present the equations as an exterior 
differential system, using $\Lie_u$ to denote the Lie derivative 
along $u$ and $\iota_u$ to denote contraction by $u$:
\begin{equation}\label{eq_Euler}
\frac{\displaystyle \del(\iota_ug)}{\displaystyle \del t} + 
	\iota_u\iota_{\curl u}\mu = -dP \;\;\;\;\; ; \;\;\;\;
	\Lie_u\mu=0.
\end{equation}
Here $P:M\ra\real^3$ is a reduced pressure function, and the 
vorticity, $\curl u$, is defined by the relation $\iota_{\curl u}\mu 
= d\iota_ug$. A vector field $u$ is said to be Eulerian if it 
satisfies Equation~(\ref{eq_Euler}) for some pressure function 
$P$. 

It follows from Bernoulli's Theorem that, for a steady Eulerian 
flow, the function $P$ is an integral of motion for flowlines.
Hence, as long as $dP$ does not vanish on open sets, steady 
Eulerian fields must be integrable. From the Fomenko-style approach 
to integrable systems, it follows that the periodic orbits of 
such a flow must have especially simple knot types (see
Theorem~\ref{thm_ZeroEnt}). 
The only alternative, then, is that $dP\equiv 0$, which translates
to the condition that $\iota_u\iota_{\curl u}\mu\equiv 0$. In 
other words, $u$ is everywhere collinear with its curl. This 
class of vector fields is of particular importance.
%%%%%%%%%%%%%%%%%%%%%%%%%%%%%%%%%%%%%%%%
\begin{dfn}
\label{def_Beltrami}
{\em
A volume-preserving vector field $u$ on a Riemannian manifold $M^3$ 
is a {\em Beltrami field} if $\curl u = f u$ for some function 
$f$ on $M$. A {\em rotational} Beltrami field is one for which
$f\neq 0$; that is, the curl is nonsingular.
}\end{dfn}
%%%%%%%%%%%%%%%%%%%%%%%%%%%%%%%%%%%%%%%%
Beltrami fields possess several interesting geometric, analytic, 
and dynamical features: see \cite{AK97,EG:I} for more information. 
A key example of a Beltrami field is the class of {\em ABC fields}:
\begin{equation}
\label{eq_ABC}
\begin{array}{l}
\dot{x} = A\sin z + C\cos y \\
\dot{y} = B\sin x + A\cos z \\
\dot{z} = C\sin y + B\cos x 
\end{array}
.\end{equation}
Here,  $A, B, C \geq 0$ are constants, and the vector field is
defined on the three-torus $T^3$. By symmetry in the variables 
and parameters, we may assume without loss of generality that 
$1=A\geq B\geq C\geq 0$. Under this convention,
the vector field is nonsingular if and only if 
$B^2 + C^2 < 1$ (see \cite{Dom+86}). 
This particular vector field exhibits the 
so-called ``Lagrangian turbulence'' --- there are apparently 
large regions of nonintegrability and mixing within the flow.

Beltrami fields occupy an important place not only within 
hydrodynamics, but also within the study of magnetic fields and 
plasmas (where they are known as {\em force-free fields}). 
As such, our results imply the existence of complex
knotting within these settings as well.

% #!#!#!#!#!#!#!#!#!#!#!#!#!#!#!#!#!#!#!#!#!#!#!#!#!#!#!#!#!
\subsection{Contact topology}
\label{sec_Contact}
% #!#!#!#!#!#!#!#!#!#!#!#!#!#!#!#!#!#!#!#!#!#!#!#!#!#!#!#!#!

We provide a brief description of the relevant concepts 
in dimension three, though the basic definitions and relationships
extend to all odd dimensions. More comprehensive treatments of this 
subject are available in \cite{MS95,Aeb94}. 

A {\em contact form} on a three-manifold $M$ is a 1-form $\alpha$ 
such $\alpha\wedge d\alpha$ is nowhere-vanishing. A {\em contact 
structure} on $M$ is the kernel of a (locally defined) contact 
form; \ie, $\xi:=\ker\alpha$. From the Frobenius integrability theorem, 
it follows that a contact structure is a totally nonintegrable plane 
field distribution on $M$. 

The interesting (and difficult) problems in contact geometry are all of 
a global nature: Darboux's Theorem (see, \eg, 
\cite{MS95,Aeb94}) implies that all contact structures are locally 
{\em contactomorphic}, or diffeomorphic preserving the plane fields.
Standard normal forms for a point include $\ker(dz+x\,dy)$ [Cartesian 
coordinates] and $\ker(dz+r^2\,d\theta)$ [polar coordinates].  
A similar result holds for a surface $\Sigma$ in a contact manifold $(M,\xi)$
as follows. Generically, $T_p\Sigma\cap \xi_p$ will be a line in $T_p\Sigma.$
This line field integrates to a singular foliation $\Sigma_\xi$ called the
{\em characteristic foliation} of $\Sigma$. The Moser-Weinstein 
Theorem implies, as in the 
single-point case of Darboux's Theorem, that $\Sigma_\xi$ determines the germ 
of $\xi$ along $\Sigma$.

Contact structures are thus implicitly global objects. Their global 
properties in dimension three depend crucially upon a dichotomy 
first explored by Bennequin \cite{Ben82} and Eliashberg \cite{Eli89}.
A contact structure $\xi$ is {\em overtwisted} if there exists an
embedded disc $D$ in $M$ whose characteristic foliation $D_\xi$ contains
a limit cycle.  If $\xi$ is not overtwisted then it is called
{\em tight}.   Eliashberg \cite{Eli89} has completely
classified overtwisted contact structures on  closed 3-manifolds --- the
geometry of overtwisted contact structures reduces to the algebra of
homotopy classes of plane fields.  Such insight into tight contact
structures is slow in coming. 

The standard contact structure on the unit $S^3\subset\real^4$ 
is given by the kernel of the 1-form 
\be
\label{eq_StdTight}
	\alpha_0 := \frac{1}{2}\left(
	x_1dx_2 - x_2dx_1 + x_3dx_4 - x_4dx_3 \right) .
\ee
The contact structure $\xi=\ker(\alpha)$ is the plane field orthogonal 
to the fibres of the Hopf fibration (orthogonal with respect to the metric 
on the unit 3-sphere induced by the standard metric on $\real^4$).
It is a foundational result that this contact structure is 
tight \cite{Ben82}; moreover, it is the unique tight structure on 
$S^3$ up to orientation and contactomorphism \cite{Eli92}.

One recently successful method for analyzing contact structures is 
to consider the dynamical structures imposed by a defining 1-form. 
Given a contact form $\alpha$, the {\em Reeb field} associated to 
$\alpha$ is the unique vector field $X$ such that 
\begin{equation}
\label{eq_Reeb}
	\iota_X\alpha = 1 \mathspace \iota_Xd\alpha = 0 .
\end{equation}
There are intricate relationships between the dynamics of Reeb fields 
and the tight/overtwisted data of the underlying contact structure 
\cite{Hof93,HWZ96b}. 

%For working with the dynamics of Reeb fields, an improved version of 
%the Darboux theorem is most helpful: 
%\begin{thm}
%\label{thm_Flowboux}
%Every contact form is locally equivalent to $dz+r^2d\theta$ [in 
%polar coordinates].
%\end{thm}
%This theorem implies that in addition to controlling the contact
%structure, one may also straighten out the Reeb field. In this respect,
%one has combined the Darboux theorem with the Flowbox theorem.
%The proof of Theorem~\ref{thm_Flowboux} appears in an inchoate form
%in the work of Giroux on convexity in contact manifolds \cite{Gir91}.
%% AH, THIS IS EASY. FIRST, USE DARBOUX TO GET ALPHA RIGHT THEN
% USE GIROUX TO STRAIGHTEN OUT THE REEB FIELD

The relationship between contact structures and solutions to the
Euler equations is explored in \cite{EG:I}, where the following 
correspondence theorem is proved:
\begin{thm}[Etnyre \& Ghrist \cite{EG:I}]
\label{thm_Correspondence}
On a fixed 3-manifold $M$, the class of vector fields which are nonsingular 
rotational Beltrami fields for some Riemannian metric $g$ and 
preserved volume form $\mu$ is equivalent 
to the class of vector fields which are nonsingular rescalings of the Reeb 
field for some contact form.
\end{thm}
In other words, given any (nonsingular, rotational) Beltrami field 
there exists a natural transverse contact form whose Reeb field 
is a reparametrization of the Beltrami field, and given any reparametrized 
Reeb field, there exists a natural Riemannian structure and volume making 
it Beltrami. For an un-rescaled Reeb field (normalized to unit length with 
respect to the contact form), the conserved volume may be chosen to be 
that induced by the Riemannian metric.

Note that the class of Beltrami fields is geometric in nature, and
is not at all well-behaved with respect to perturbations, etc. 
On the other hand, the Reeb fields and their rescalings are quite
flexible: a fact we shall take advantage of in 
\S\ref{sec_Proof}. Shedding the metric constraints 
thus transforms geometric problems to topological ones.

% #!#!#!#!#!#!#!#!#!#!#!#!#!#!#!#!#!#!#!#!#!#!#!#!#!#!#!#!#!
\subsection{Template theory}
\label{sec_Template}
% #!#!#!#!#!#!#!#!#!#!#!#!#!#!#!#!#!#!#!#!#!#!#!#!#!#!#!#!#!

For a complete treatment of this subject, see \cite{GHS97}.

The problem of knotted orbits in vector fields on three-manifolds 
is full of surprises, beginning with the pioneering work of 
Williams in the late 1970's to understand solutions to the 
Lorenz equations \cite{Wil77}. In developing the geometric model
for the Lorenz equations \cite{GW79}, Williams considered 
branched surfaces. In a pair of papers with Birman \cite{BW83a,BW83b}, 
the knot-theoretical implications of these ideas were brought 
forth in the notion of a {\em knotholder}, later rechristened 
a {\em template} \cite{HW85}.
\begin{dfn}{\em
A {\em template} is a compact branched 2-manifold with boundary 
supporting a smooth expansive semiflow.
}\end{dfn}
%

% ??????????????? FIGURE CHARTS ?????????????????
\begin{figure}[htb]
\begin{center}
\begin{psfrags}
\psfrag{a}[][]{(a)}
\psfrag{b}[][]{(b)}
	\epsfxsize=3.5in\leavevmode\epsfbox{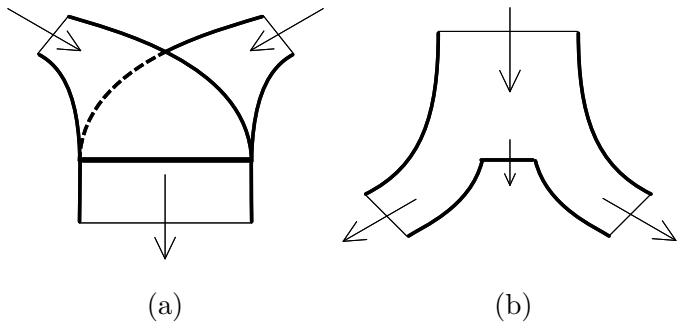}
\end{psfrags}
\end{center}
\caption{(a) Joining and (b) splitting charts for templates.}
\label{fig_Charts}
\end{figure}
\begin{figure}[hbt]
\begin{center}
\begin{psfrags}
        \psfrag{a}[][]{(a)}
        \psfrag{b}[][]{(b)}
	\psfrag{c}[][]{(c)}
	\epsfxsize=5.0in\leavevmode\epsfbox{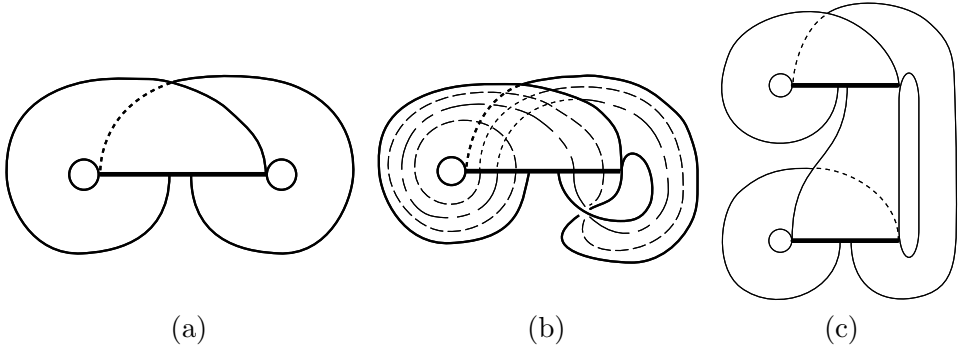}
\end{psfrags}
\end{center}
\caption{(a) The Lorenz template; (b) A template ($\L(0,1)$) 
	with a pair of closed orbits; (c) The universal template $\V$.}
\label{fig_Example}
\end{figure}

% ????????????????????????????????????????????????????
Templates have a description in terms of charts: every template
is diffeomorphic to the branched surface obtained by gluing 
together a finite number of {\em joining} and {\em splitting 
charts} (illustrated in Figure~\ref{fig_Charts}) end-to-end 
respecting the semiflows. Examples of embedded templates appear
in Figure~\ref{fig_Example}.

% ????????????? FIGURE EXAMPLE ???????????
% ??????????????????????????????????????????????

The semiflow on a template is ``overflowing'' in the sense that
it is not defined on the gaps between strips of the splitting charts. 
This is inconsequential as one is only interested in the invariant
sets of the semiflow; hence, the gaps are often back-propagated to 
the branchlines in figures. Due to the expanding nature of the 
dynamics on a template, the invariant set consists of the 
suspension of a subshift of finite type, generated by the symbolic 
dynamics of the Markov partition ${\cal R}$ induced by the branch 
lines. Specifically, the branchlines are partitioned by the 
exit sets into a finite number of intervals $\{R_1,\ldots,R_n\}$,
each of which gets stretched under the semiflow to cover two 
other partition elements. The expansive nature of the semiflow
implies that forward orbits on the template are in bijective 
correspondence with the semi-infinite admissible symbol 
sequences in the alphabet induced by ${\cal R}$ --- see 
\cite{GHS97} for more information and examples.

Templates arise naturally in the context of nontrivial hyperbolic 
invariant sets in a flow on a three-manifold $M$. Let $\Lambda$ denote
such an invariant set. The Template Theorem of Birman and Williams
implies that there exists an embedded template $\T_\Lambda\subset M$
such that the periodic orbits of $\Lambda$ are in bijective 
correspondence with those of the semiflow on $\T_\Lambda$, and 
that, furthermore, this correspondence preserves all knotting 
and linking information. Hence, to obtain information about 
knotted periodic orbits in $M$, one simply analyzes the 
template $\T_\Lambda$. The essence of their proof is to collapse 
a foliation by strong stable manifolds --- identifying all orbits
with the same asymptotic future. Clearly, this preserves the 
periodic orbit set and its embedding properties. 

Given the Template Theorem above, one may proceed to analyze the 
knotting and linking properties of various systems. This analysis has 
been conducted for the (geometric) Lorenz attractor \cite{BW83a},
the suspension of the Smale horseshoe \cite{BW83b,HW85,Hol86,GH93},
systems associated with a Josephson junction \cite{Hol87}, 
flows near Shilnikov homoclinic orbits \cite{GH96}, and 
flows transverse to fibred links in $S^3$ with pseudo-Anosov
monodromy \cite{BW83b,G97TOP}. 

The question of whether a flow on $S^3$ can support all knots 
at once may thus be addressed from the point of view of templates. 
In \cite{G97TOP} it was shown that there exist {\em universal 
templates} in $S^3$ which contain closed orbits of all possible 
knot (and link) types. The canonical example appears in 
Figure~\ref{fig_Example}(c). The way in which the knots lie
within the template is highly nontrivial: the simplest known 
example of a figure-eight knot on this template crosses the 
branchlines millions of times \cite{G97TOP,GHS97}. 

\begin{figure}[htb]
\begin{center}
\begin{psfrags}
	\psfrag{m}[b][]{$m$}
	\psfrag{n}[b][]{$n$}
%	\psfrag{a}[][]{(a)}
%	\psfrag{b}[][]{(b)}
%	\psfrag{c}[][]{(c)}
	\epsfxsize=5.1in\leavevmode\epsfbox{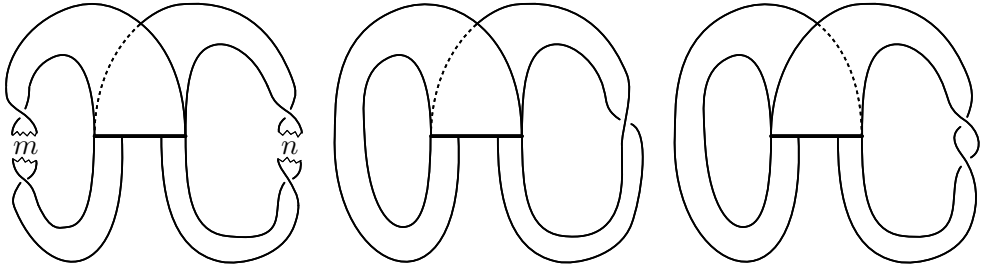}
\end{psfrags}
\end{center}
\caption{[left] The Lorenz-like template $\L(m,n)$; [center] $\L(0,-1)$; 
	[right] $\L^*(0,2)$. Note that positive coefficients imply 
	left-handed twists.}
\label{fig_LorenzLike}
\end{figure}

Given a flow on $S^3$ which supports a hyperbolic invariant 
set modeled by a template, it is in general impossible to determine 
if this template is universal: no general computable criterion
is known. However, the only obstruction to being universal is
on the embedding level --- any (abstract) template can be 
embedded in $S^3$ so as to be universal \cite[Thm. 3.3.5]{GHS97}. 
A particularly useful result concerns the {\em Lorenz-like
templates}, $\L(m,n)$, pictured in Figure~\ref{fig_LorenzLike}. 
Also useful in the sequel are the Lorenz-like templates with 
the branchline crossing reversed: denote these by $\L^*(m,n)$, 
with the sign convention as in Figure~\ref{fig_LorenzLike}.
% ??????????????? FIGURE LORENZLIKE ?????????????????
% ????????????????????????????????????????????????????

\begin{thm}[Ghrist \cite{G97TOP}]
\label{thm_Lorenzlike}
In the case where the product $mn\geq 0$, the Lorenz-like template
$\L(m,n)$ is universal if and only if $mn=0$ and $m+n<0$.
\end{thm}

% **************************************
\section{A Reeb field with all knots}
\label{sec_Proof}
% **************************************

In this section, we prove the existence of a Reeb field on the 
tight 3-sphere whose flow has a hyperbolic invariant subset which 
collapses to a universal template. 
We begin by noting the necessity of the contact-topological techniques
outlined in \S\ref{sec_Beltrami}. Recall that
the class of {\em zero-entropy knots} consists of those 
knots obtainable from the unknot by iterating the operations of 
cabling and connected sum \cite{EG:unknot}. Such knots are a very small
subclass of knots, excluding such large classes as hyperbolic 
knots (knots whose complement in $S^3$ supports a hyperbolic 
geometry). 

\begin{thm}[Etnyre \& Ghrist \cite{EG:unknot}]
\label{thm_ZeroEnt}
Let $u$ denote a $C^\omega$ steady nonsingular solution to the 
Euler equations on a Riemannian 3-manifold. If $u$ is not 
a Beltrami field, then every periodic orbit of $u$ must be a 
{\em zero-entropy} knot. 
\end{thm}

Hence we conclude that it is necessary to consider Reeb (\ie, Beltrami)
fields as the only possibility for constructing highly regular 
steady flows with all knots. 

\begin{lemma}
\label{lem_Some}
There exists a Reeb field on some tight contact 3-manifold possessing
a nontrivial one-dimensional hyperbolic invariant set.
\end{lemma}
\pf
We give an explicit example on the 3-torus.
Consider the ABC equations of \S\ref{sec_Beltrami}.
From Equations~(\ref{eq_ABC}) and (\ref{eq_Reeb}) it follows 
that the ABC fields lie within the kernel of the derivative 
of the 1-form 
\begin{equation}
\alpha:=(A\sin z + C\cos y)dx+(B\sin x + A\cos z)dy+(C\sin y + B\cos x)dz ,
\end{equation}
and that this is a contact form when the vector field is nonsingular. 
Denote by $\xi:=\ker(\alpha)$ the induced contact structure on $T^3$.
This contact structure is always tight \cite{EG:I}.

In the limit where $A=1, B=1/2$, and $C=0$, the vector field 
takes on the particularly simple form 
\be
\begin{array}{l}
	\dot x = \sin z \\
	\dot y = \frac{1}{2}\sin x + \cos z \\
	\dot z = \frac{1}{2}\cos x 
\end{array} .
\ee
It is straightforward to demonstrate that there exists a pair of periodic 
orbits whose stable and unstable invariant manifolds intersect each
other nontransversally (see, \eg, \cite{Dom+86}). Upon
perturbing $C$ to a small nonzero value, this connection may become
transverse. Indeed, a Melnikov perturbation analysis reveals
precisely this fact \cite{HZD98,ZKBH93,Gau85}.
It thus follows from the Birkhoff-Smale Homoclinic Theorem 
that there exists parameters for which Equation~(\ref{eq_ABC})
possesses a nontrivial 1-d hyperbolic invariant set as a solution: 
a suspended 2-shift. 
\qed

%-----------------------------------------------------------------------

At this stage, there are two possible ways to proceed. One could 
perform a straightforward surgery construction on a tubular neighborhood of 
the hyperbolic 2-shift above to obtain a Reeb field on $S^3$ 
having an invariant set modeled by a Lorenz-like template. However, 
it is not at all obvious that the contact structure so induced on 
$S^3$ is the tight one, especially under the constraint that the 
framing on the surgery coefficients be such that the resulting 
template is universal. Thus, we turn to a method of constructing 
a contact embedding into the tight 3-sphere handle-by-handle.

In order to embed this hyperbolic 2-shift into the tight three-sphere,
we require the following technical lemma for controlling characteristic
foliations on annuli. This Lemma has appeared in the preprint 
\cite[Lemma 3.3]{Col99}, as well as in \cite[Lemma 4.4]{Mak98} in 
a slightly more restricted setting. We include the simple 
proof for completeness and clarity.
\begin{lemma}
\label{lem_Monodromy}
Given any orientation-preserving diffeomorphism $f:S^1\ra S^1$, 
there exists a smooth annulus $A$ in $\real^3$ such that 
(1) $A$ is transverse to the 
contact structure $\lambda:=dz+r^2d\theta$; (2) $\del A$ consists of the 
circles $\{r=\epsilon; z=\pm 1\}$; and (3) the monodromy obtained by 
sliding along leaves of $A_\lambda$ from $z=-1$ to $z=+1$ is precisely 
the map $f$. 
\end{lemma}
\pf
Begin with the annulus $\{r=\epsilon; z\in[-1,1]\}\subset\real^3$ 
and outfit it with any foliation ${\cal F}$ such that (1) the 
slope of the leaves of ${\cal F}$ is always negative; and (2) the monodromy
along ${\cal F}$ exists and is given by $f$. Such a foliation clearly
exists. To realize ${\cal F}$ 
as the characteristic foliation of an annulus in $\real^3$, 
simply compute the slope $-g(\theta,z)$ of the leaves of ${\cal F}$
in $(\theta,z)$ coordinates. Then, the annulus 
$A:=\{(\sqrt{g(\theta,z)},\theta,z) ; \theta\in S^1, z\in[-1,1]\}$ 
has characteristic foliation given by ${\cal F}$, since $A_\lambda$ is 
defined by the relation $\frac{dz}{d\theta}=-r^2$. To fix the 
boundary of $A$, simply choose ${\cal F}$ so that the slope at
$z=\pm 1$ is precisely $-\epsilon^2$: this in no ways hinders the 
choice of monodromy. 
\qed

%-----------------------------------------------------------------------
\begin{thm}
\label{thm_Embed}
There exists a hyperbolic invariant suspended 2-shift $\La\subset T^3$ 
in the Reeb field of Lemma~\ref{lem_Some} and a tubular 
neighborhood of $\Lambda$, $N$, diffeomorphic to a genus-2 
handlebody such that $(N,\xi)$ embeds contactomorphically 
into $(S^3,\xi_0)$. 
\end{thm}
\pf
Choose $\gamma_1$ a closed orbit in the hyperbolic invariant 
set from Lemma~\ref{lem_Some} and $\kappa_1$ 
an unknotted curve transverse to $\xi_0$ in $S^3$. By the
Moser-Weinstein Theorem, there exists a neighborhood $N_1$ 
of $\gamma_1$ and a contact embedding
$\Phi:N_1\hra S^3$ taking $\gamma_1$ to $\kappa_1$. 

Let $\Sigma$ denote a small disc transverse to the flow at a point
$p_1\in\gamma_1$. It is a standard result from the theory of 
hyperbolic dynamics that the periodic orbits are dense in the 
invariant 2-shift; thus, choose $\gamma_2$ a closed orbit in the 
invariant set intersecting $\Sigma$ once in the point $p_2$: such 
an orbit exists for $\Sigma$ sufficiently small. 

\begin{figure}[htb]
\begin{center}
\begin{psfrags}
        \psfrag{s}[br][br]{$\Sigma$}
        \psfrag{U}[][]{$U$}
	\psfrag{a}[br][]{$H_1$}
	\psfrag{b}[bl][bl]{$H_2$}
	\psfrag{c}[][]{$R_1$}
	\psfrag{d}[b][]{$R_2$}
 	\epsfxsize=3.25in\leavevmode\epsfbox{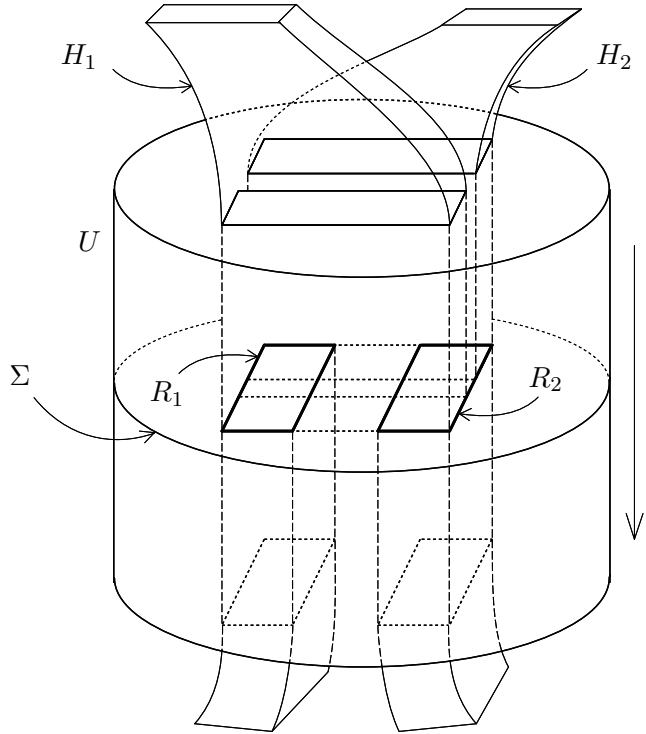}
\end{psfrags}
\end{center}
\caption{The neighborhood $U$ in $T^3$. The 1-handles $H_1$ and
$H_2$ (drawn without rounded corners as the forward images of the $R_i$) 
form a genus-2 handlebody.}
\label{fig_Flowbox}
\end{figure}

There exists a ``small'' suspended 2-shift $\La$ that is generated
by $\gamma_1$ and $\gamma_2$ as follows. 
There exists a Markov partition ${R_1,R_2}\subset\Sigma$ 
by a pair of rectangles on $\Sigma$, such that the 
fixed points of the return map on $R_1\cup R_2$ consists of
the pair of points $p_1$ and $p_2$. 
From basic symbolic dynamics it follows that the Poincar\'e 
return map restricted to $R_1\cup R_2$ possesses a hyperbolic 
invariant 2-shift. Let $U$ denote a flowbox neighborhood of 
$\Sigma$, and let $H_1$ and $H_2$ denote, respectively, the solid cylinders
in the complement of $U$ traced out by a small neighborhood of 
$R_1$ and $R_2$ (resp.) under the flow, as illustrated in 
Figure~\ref{fig_Flowbox}. The union $N:=U\cup H_1\cup H_2$ 
is thus a genus-2 handlebody neighborhood of $\La$. 

% ??????????????? FIGURE FLOWBOX ?????????????????
% ????????????????????????????????????????????????????

Note that the contactomorphism $\Phi$ can be defined on $U$ and $H_1$, since
the size of the Markov rectangles is bounded by the size of $\Sigma$ chosen
arbitrarily small. Once this is fixed, however, the size of the 
neighborhood $H_2$ of $\gamma_2$ cannot then be chosen to be
arbitrarily small, while still containing an invariant 2-shift. Nor 
can we fix $\gamma_1$ and $\gamma_2$ and then choose arbitrarily 
small neighborhoods; hence, one is impeded from employing a
Moser-Weinstein type argument to extend $\Phi$ to $H_2$. We circumvent this
by using the rigidity of the tight contact structures implicated
in the construction. The remainder of this proof is rather technical ---
we organize the more intricate steps in a series of claims.

Denote by $A$ the smooth annulus $\del H_2$ 
with boundary components $A^+$ and $A^-$ on $\del U$. 
Let $B^+$ and $B^-$ denote the $\Phi$-images of
$A^+$, and $A^-$ respectively. Denote by $\kappa_0\subset\Phi(U)$ 
the image of $\gamma_2\cap U$. This curve connects the two 
disc-components $D^\pm$ of $\del(\Phi(U))-B$. 
Choose $\kappa'\subset S^3-\Phi(N_1\cup U)$ an arc in $S^3$ 
transverse to $\xi$ connecting the ends of $\kappa_0$ such that 
$\kappa_2:=\kappa_0\cup\kappa'$ is a smooth simple closed curve which 
bounds an embedded disc in the complement of $\kappa_1$. 
This curve will become the core of the image of $H_2$ in $S^3$.

{\bf Claim 1:} {\em There exists a cylinder $B$ embedded in $S^3$
connecting $B^-$ to $B^+$ such that $B$ is everywhere transverse
to $\xi_0$.}

{\em Proof of Claim 1:} 
Let $B'$ be the boundary of a small tubular neighborhood of 
$\kappa'$ contactomorphic to $\{(r,\theta, z) : r\leq\epsilon; \, z=\pm 2\}$
with the contact structure $dz+r^2d\theta$ from Lemma~\ref{lem_Monodromy}. 
We may also assume that the discs $D'_\pm$ corresponding 
to $z=\pm 2$ are properly contained in the discs 
$D^\pm$ bounded by $B^\pm$. We require the following:

{\bf Claim 2:} {\em
After a modification of the flowbox $U$, the characteristic 
foliations on $D^\pm$ contain exactly one elliptic singularity 
which occurs at $\kappa'\cap D^\pm$.}

Assuming Claim 2, it is clear that the characteristic foliations on the 
annuli $D^\pm-D'_\pm$ are by radial lines. Hence 
we may use these annuli to drag the two circles $\partial B'$ 
to $B^\pm$ without introducing any singularities in the 
characteristic foliation. The resulting annulus is our desired $B$.
\qed$_{1}$.

{\em Proof of Claim 2:}
We begin by carefully choosing $U$. Our original choice of
$\Sigma$ can be taken to be a small disc with one elliptic singularity 
in the center, since the disc is transverse to the Reeb field and 
hence all tangencies between the [oriented] disc and 
the [oriented] plane field must have the same signs, 
allowing one to cancel the singularities (as in, \eg, \cite{Eli92}). 
We may then use the Reeb flow to 
construct a contactomorphism from a neighborhood $U'$ of $\Sigma$ 
to a neighborhood of the origin in $\real^3$ 
(with polar coordinates and the standard contact structure) 
taking the Reeb field arbitrarily close to $\frac{\del}{\del z}$. 
One may now assume that the entire construction takes place within 
this local model. As such, $\gamma_2$ may be chosen so that 
the rectangle $R_2$ is of sufficiently small diameter.
From this it follows that the leaves of the characteristic foliation on the
boundary of the tube $T$ generated by $R_2$ under
the Reeb flow wrap many times around $T$. Hence 
we may easily choose a curve $c$ on $\partial T$
that is isotopic to $\del R_2$ but transverse 
to $\xi_0$. The curve $c$ bounds a disc in $T$ with precisely one 
elliptic singularity (this can be arranged as before 
since the Reeb field is transverse to the disc). 
We may now flow this disc forwards by the Reeb flow
[preserving the characteristic foliations]
so that we have two copies of it near $\del U'$, and 
we may finally isotope $U'$ so that its
boundary contains these copies. This is the desired $U$.
\qed$_2$

{\bf Claim 3:} {\em
The embedding $\Phi$ extends to a contactomorphism
from a neighborhood of $A$ to a neighborhood of $B$. }

{\em Proof of Claim 3:}
Consider the characteristic foliation $A_{\xi}$: it is nonsingular
since the Reeb field is tangent to $A$. Furthermore, there are no 
meridional closed curves in $A_{\xi}$ by tightness.
%since, in the proof of Claim 2, we constructed
%a transverse meridional curve propagated by the Reeb field along $A$.
Thus, the monodromy map given by sliding along leaves of $A_\xi$
exists. The same is true for $B_{\xi_0}$ --- this 
follows from Claims 1 and 2. Since away from $U$ the 
annulus $B$ is contactomorphic to the cylinder $\{r=\epsilon, 
z=\pm 1\}$ in $(\real^3,dz+r^2d\theta)$, we may apply 
Lemma~\ref{lem_Monodromy} above to modify $B$ 
rel the planes $z=\pm 1$ so that the characteristic foliation 
$B_{\xi_0}$ realizes the same monodromy as $A_\xi$. Thus $\Phi$ 
extends to a diffeomorphism which takes the characteristic foliation 
of $A$ to the that of $B$. The Moser-Weinstein Theorem completes
the proof of the claim.
\qed$_3$

To complete the proof of Theorem~\ref{thm_Embed},
cap off the ends of the cylinder $A$ by a pair of discs within $U$, 
forming a smooth 2-sphere $\overline{A}$. The $\Phi$-images of these 
discs cap off $B$ to a sphere $\overline{B}$. 
By Lemma~\ref{lem_Some}, the structure $\xi$ on 
$T^3$ is tight; thus, we have tight solid balls whose boundaries are 
contactomorphic. A celebrated theorem of Eliashberg \cite{Eli92} 
states that any tight 3-balls with the same characteristic foliations 
on the boundaries are in fact contact isotopic rel the boundaries. 
Hence, we may extend $\Phi$ to a contact embedding of 
$N=H_1\cup H_2\cup U\subset T^3$ into $(S^3,\xi_0)$.
\qed
%-----------------------------------------------------------------------

The resulting Reeb field on $\Phi(N)$ given by the contact form 
$\Phi_*\alpha$ has an invariant set whose
template is a Lorenz-like template of type $\L(m,n)$ (or $\L^*(m,n)$)
for some integers $m,n\in\zed$. We must control the  
twisting in order to apply Theorem~\ref{thm_Lorenzlike}.
This we do by changing the embedding along $H_1$ and $H_2$ to include
extra meridional twists. However, due to the implicit handedness in 
a contact structure, it is possible to make arbitrary twists 
on $H_2$ in only one direction, as shown below.

%-----------------------------------------------------------------------
\begin{prop}
\label{prop_Twist}
The contact embedding $\Phi:N\hra S^3$ can be chosen so that
the invariant set $\Phi(\La)$ is modeled by a universal template.
\end{prop}
\pf
We show how to manipulate the embedding $\Phi$ so that the image of 
$\La$ in $S^3$ is modeled by a Lorenz-like template of type $\L(0,-n)$
(or its mirror image), then apply Theorem~\ref{thm_Lorenzlike}.
First we control the embedding of $H_1$ into $S^3$ to obtain 
zero twist. We recapitulate the initial steps of Theorem~\ref{thm_Embed},
in particular the embedding of a neighborhood of $\gamma_1$ into $S^3$. 
After fixing a framing for the normal bundles of $\gamma_1$ and 
$\kappa_1$ respectively, 
there are an integer's worth of choices of isotopy classes of 
embeddings, depending upon the twist of the normal bundle 
(with respect to the prescribed framings). Changing the isotopy 
class has the effect of modifying the twist associated to the
local (2-d) stable manifold of $\kappa_1$ in $S^3$ under the 
Reeb field of $\Phi_*(\alpha)$. Any such isotopy class may
be realized by a contact embedding as follows. Clearly an embedding may be 
chosen which takes $\rest{\xi}{\gamma_1}$ to the corresponding 
planes of $\rest{\xi_0}{\kappa_1}$ in $S^3$. 
Then, the Moser-Weinstein Theorem 
implies that this extends to a contact embedding on a tubular
neighborhood. Thus, choose a contact embedding on a neighborhood
of $\gamma_1$ which sends it to a curve whose local stable
manifold has zero-twist, yielding a
Lorenz-like template of type $\L(0,n)$ or $\L^*(0,n)$ for some $n$. 

To modify the twist $n$ on the image of $H_2$, note that $H_2$ is not 
an arbitrarily small neighborhood of $\gamma_2$, so the preceding argument 
is invalid. However, one can modify the number of twists on the 
image of $H_2$ by choosing the annulus $B$ carefully. In the 
construction of Theorem~\ref{thm_Embed}, the crucial step is to 
have the monodromy on $B$ agree with that of the tube $A$ in 
$T^3$. From the proof of Lemma~\ref{lem_Monodromy}, it is clear 
that one can choose thinner and thinner annuli for $B$ which maintain 
the monodromy, but which increment the twisting in the characteristic 
foliation $B_{\xi_0}$ by full left-handed twists. Thus, the effect 
of modifying $B$ to  the ``next'' smaller tube changes the 
associated template from $\L(0,n)$ to $\L(0,n+1)$ (or from $\L^*(0,n)$ 
to $\L^*(0,n+1)$). Decreasing the value of $n$ would be 
possible only if one can increase the size of the tube bounded by 
$B$ in $S^3$ --- this is in general impossible. Thus, only templates 
of the form $\L(0,n)$ or $\L^*(0,n)$ for 
$n$ an arbitrarily large positive integer may be constructed.

\begin{figure}[htb]
\begin{center}
\begin{psfrags}
        \psfrag{1}[][]{$\gamma_1$}
        \psfrag{2}[][]{$\gamma_2$}
	\epsfxsize=1.6in\leavevmode\epsfbox{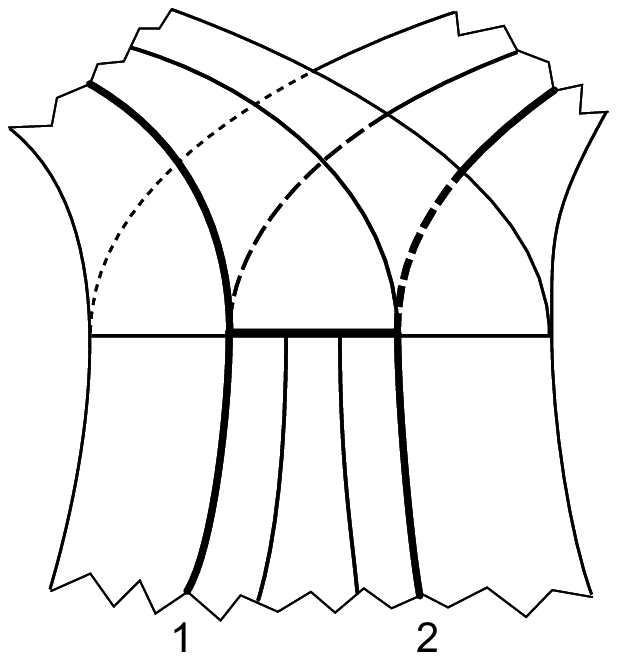}
	\epsfxsize=1.6in\leavevmode\epsfbox{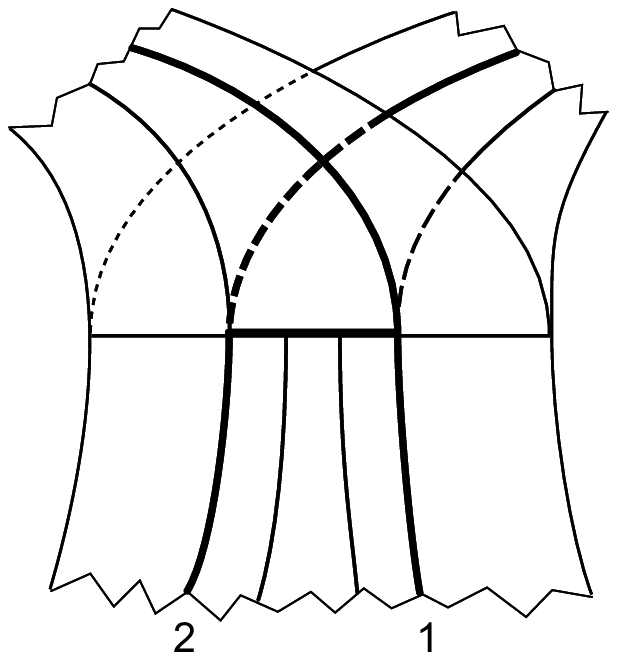}
	\epsfxsize=1.6in\leavevmode\epsfbox{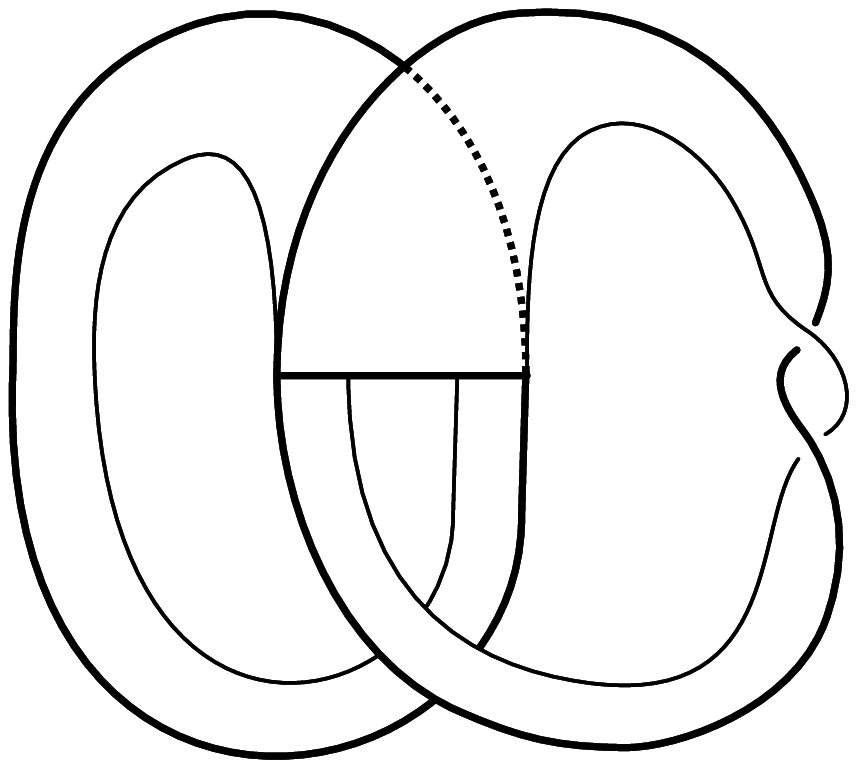}
\end{psfrags}
\end{center}
\caption{One may choose $\gamma_2$ to the right of $\gamma_1$ [left] 
or to the left of $\gamma_1$ [center]. In the case of a left-over-right
branchline crossing, the latter choice creates a template [right]
which is isotopic to $\L^*(0,n)$. (Flip the branchline over to put 
this template into normal form.)}
\label{fig_Switch}
\end{figure}

For such an $n$, the template $\L(0,n)$ is definitely not universal,
whereas $\L^*(0,n)$, being the mirror image of $\L(0,-n)$, is 
universal by Theorem~\ref{thm_Lorenzlike}. We must thus control the 
sign of the crossing of the strips ``at the branchline'' (\cf
Figure~\ref{fig_LorenzLike}). This
sign is determined by the choice of $\gamma_2$ in $T^3$
as follows. Having fixed the neighborhood $U$, it is well-known 
that there is a local product structure on $\rest{\La}{U}$ by the 
stable and unstable manifolds of $\La$. The template is obtained 
by collapsing out the local strong stable foliation, thus 
determining a regular projection for the template. The curves
$\gamma_1$ and $\gamma_2$ belong to separate branchline strips
whose crossing sign is thus fixed. In the case where the 
natural crossing sign yields $\L(0,n)$, we choose a different 
curve for $\gamma_2$ and repeat the
entire construction. Figure~\ref{fig_Switch} illustrates that 
choosing $\gamma_2$ to lie ``to the left'' of $\gamma_1$ under the 
canonical projection switches the crossing at the branchline when 
the image template in $S^3$ is isotoped to normal form. Hence, the 
resulting template is isotopic to the universal template $\L^*(0,n)$.
\qed

% ??????????????? FIGURE SWITCH ?????????????????
% ????????????????????????????????????????????????????

%----------------------------------------------------------------------- 
\begin{thm}
\label{thm_Hard}
There exists a tight contact form on $S^3$ whose Reeb field 
possesses periodic orbits of all possible knot and link types
simultaneously.
\end{thm}
\pf
We have constructed a contact embedding $\Phi:N\hra S^3$ from the 
genus-two handlebody to tight $S^3$. Pushing forward the 
form $\alpha$ on $T^3$ via $\Phi$ yields $f\alpha_0$ on 
$\Phi(N)\subset S^3$, where $\alpha_0$ denotes the standard tight 
contact form of Equation~(\ref{eq_StdTight}) and $f>0$. 
By extending $f$ smoothly to a positive function on all of $S^3$, one has 
a tight contact form $f\alpha_0$, yielding the desired Reeb field.
\qed
\vspace{0.1in}

\noindent
{\bf Remark:}
The problem of classifying knots and links which are everywhere 
transverse to (or tangent to) a tight contact structure up to 
isotopy within said class is particularly delicate \cite{Eli93}.
We have demonstrated that there are no obstructions to the 
existence of all topological knot and link types simultaneously as 
transverse knots. However, it is not true that all transverse knot
types are realized in our constructions --- there are natural 
restrictions due to tightness. Beyond this, it is highly unlikely
that all tight transverse knot types are present in the 
flows constructed here. From \cite[Ch. 3]{GHS97}, it follows that
even simple knot types have only very complicated presentations on
universal templates. A simple calculation implies that the 
self-linking numbers (an invariant of transverse knot types) of 
simple knots on a universal template may be 
astronomically large negative numbers. It would be interesting 
to see exactly how such self-linking numbers are distributed, as
well as if it is possible to control the self-linking numbers.

\begin{cor}
\label{cor_Hard}
There exists a steady nonsingular $C^\omega$ solution to the 
Euler equations on a Riemannian $S^3$ which possesses 
flowlines of all possible knot and link types simultaneously.
\end{cor}
\pf
The contact form assembled in Theorem~\ref{thm_Hard} is $f\alpha_0$, 
where $f$ is a positive function and $\alpha_0$ is the standard tight 
form on $S^3$. As $\alpha_0$ is an analytic form, we may construct an
analytic contact form by perturbing $f$ to be $C^\omega$. By the 
structural stability of hyperbolic invariant sets, 
one has that the invariant set modeled by the universal template 
persists under the perturbation. 
The correspondence of Theorem~\ref{thm_Correspondence} 
completes the proof.
\qed

The natural question with which we conclude is:

\vspace{0.1in}

\noindent
{\bf Question:} {\em Are there steady solutions to the Euler
equations having all knots and links for the Euclidean 
metric on $\real^3$ (or for the round metric on $S^3$)?}

\small
% BIBLIOGRAPHY
%\bibliography{../../refs/references}
\bibliographystyle{alpha}

\newcommand{\etalchar}[1]{$^{#1}$}

\end{document}